# Machine Learning Applications in the Routing in Computer Networks

**Ke Liang**
*Pennsylvania State University*
*Dept. of CSE*
*kul660@psu.edu*

**Mitchel Myers**
*Pennsylvania State University*
*Dept. of CSE*
*mtm387@psu.edu*

## 1 Introduction to the topic

With the rapid development of the Internet in many areas, higher speed and larger capacity of the network for new Internet applications are required. However, some of the traditional network technologies meet bottlenecks when it comes to satisfying the increasing demand. Meanwhile, machine learning technologies have shown great potential in many regions in recent years. Based on this situation, it is worthwhile to research applications of machine learning in the computer network field.

Network Routing, which determines the route taken by packets from source to destination [1], plays a critical role in networking for selecting the path for packet transmission. Different operation policies and objectives determine which selection criteria we need for choosing or designing the routing method. Cost minimization, maximization of link utilization, and QoS provisioning are all common criteria [2]. Normally, traditional routing algorithms are designed based on one criterion, which do not scale well for different types of real-world scenarios. For example, as shown in [3], with Open Shortest Path First (OSPF), all the packets will be delivered via the 100 Mbps path, and this will waste resources along the other links and lead to network congestion. We desired development of new algorithms which are better at utilizing the network resources.

For many years, researchers have devoted themselves to designing routing algorithms for better reachability, performance, and scalability. One of the directions is the combination of machine learning technologies and routing algorithms.

Machine learning can be divided into 3 classes, supervised machine learning, unsupervised machine learning, and reinforcement learning based on the popular connectionism [4-5]. Each of these classes is quickly developing and evolving in many areas, including medical image processing, computer vision, and natural language processing. One of their advantages is the huge potential for improving the scalability for the model, due to the data driven mechanism. Especially with the development of graph neural networks and reinforcement learning, there is much research being done in that direction, mainly because computer network scenarios can be easily described with the graph, and reinforcement learning can be easily applied for the objective value. The machine learning technologies can be used in both centralized routing algorithms and decentralized routing algorithms. For each of the classes, multiple machine learning methods are used in two ways, on-line and off-line.

We aim to give a relatively comprehensive survey for introducing some novel machine learning applications in routing in computer networks. The assumption of all the methods in the paper is that they are applications combining with routing and supervised learning or reinforcement learning technologies. In section 2, we explain how and why we classify the literature. In section 3, we give the details, pros, and cons of each paper for each class. In section 4, we implement and evaluate two approaches, and we explain the results and interpretations for each method and experiments. In section 5, we summarize the main results of the paper, together with some discussions on emerging trends. Section 6 shows the contributions of each team member.

## 2   Classification

Routing algorithms can be divided into 2 classes: centralized methods and decentralized methods. In centralized methods, a logically centralized controller is responsible for observing the network and making routing decisions. In a decentralized architecture, such as traditional networking, each router decides how to route its packets, but may exchange messages with other nodes to learn about the rest of the network. In this paper, we classify the literatures into these two classes, but we also further subdivide each class into two subclasses: supervised learning methods (e.g. Convolutional Neural Networks, Graph Neural Networks, and Recurrent Neural Networks, etc.) and reinforcement learning methods (e.g. Q-Learning, Deep Q-Learning, etc.). The reasons for these two subclasses are based on classifications of the machine learning methods from the popular connectionism as mentioned in introduction, which is the most popular theory for machine learning these years [4-5]. Generally speaking, the idea of connectionism is from neural biology. It treats the unit of the machine learning model as the neurons. More details for each class will be introduced in section 3.

## 3   Presentation of each class

This section will be divided into 3 parts, including centralized routing algorithms, decentralized routing algorithms, and comparison of each class. In the first two subsections, the details of each class are explained based on the type of machine learning methods, including supervised learning methods and reinforcement learning methods. For each subclass, the literatures are introduced based on the time. In the third part of the section, we make a comparison table to conclude the methods described.

### 3.1   Centralized Routing Algorithms

#### 3.1.1 Supervised Learning Methods

In 2018, Zirui Zhuang, et al. [6] proposed the Graph-Aware Deep Learning based intelligent routing strategy (GADL), which aimed to solve the problems, such as slow convergence and performance decline when configuring the complex dynamics condition in networking. The GADL consisted of 2 parts, including the graph kernel and the convolutional neural network, and they also used a feature processing model to convert the network state measurements into the representative features. The phases of the method include tracking of the network state measurements continuously, updating the optimal path, calculating the optimal path via machine learning method (as shown in [6]). The experiments showed that GADL had better learning performance than the previous CNN and DBA models, and the average network latency of it was lower than OSPF (as shown in [6]) and was more stable than DBA.

In 2019, Gangxiang Shen, et al. [7] proposed the ML-assisted Least Loaded (LL) algorithm which was the combination of the supervised naïve Bayes classifier and the original LL algorithm to conquer the connection failure problems within the following situation in circuit-switched networks. The structure of the ML-assisted Least Loaded (LL) algorithm is proposed in [7]. When a request was received, all possible paths between the nodes were considered. For those which are not already at capacity, the algorithm computed the current load for each path. It also considered the state of the network if a connection was established over the path, hypothetically updating the loads. Then, the naïve Bayes classifier was used to determine how likely a blocked connection would be in the new network state. The candidate paths were scored using both the current load and future blocking chance. The algorithm established the connection over the path with the best overall score and constantly updated its parameters in response to network events. The experiments were carried out in an SDN-like fashion with a central control in charge of evaluating and establishing paths, which showed the ML-assisted LL algorithm outperforms the LL algorithm significantly in NSFNET and ARPA-2 datasets. The results are shown in [7].

In 2019 to 2020, K. Rusek, et al. [8-9] proposed RouteNet, which was a novel network model based on graph neural networks (GNNs). In this method, machine learning was used to help optimize the routing scheme based for the average delay. The model made use of the ability of GNNs to learn and model the graph-based information.

It first came up with a way to model the program. It used a set of the links $N = \{l_i \mid i \in (1, \ldots, n_l)\}$ to represent the network and the routing scheme could be regarded as by a set of paths $P = \{p_k \mid k \in (1, \ldots, n_p)\}$, where each path consisted of a set of sequential links $p_k = (l_k(1), \ldots, l_k(|p_k|))$ (and $k(i)$ is the index of the $i$-th link in path $k$). GNNs models usually have three basic functions: message, update, and readout. As for the RouteNet, the update functions are based on the input routing scheme. The architecture of the RouteNet is flexible to represent any source-destination routing scheme, and the Recurrent Neural Network (RNN) is used to aggregate link states on paths. The authors test the model with the QoS-aware routing optimization scenario and achieve good results for both mean delay and packet loss compared to the traditional shortest path methods.

### 3.1.2 Reinforcement Learning Methods

In 2016, S. Lin et al. [10] proposed the QoS-aware adaptive reinforcement learning routing method in multi-layer hierarchical software defined networks. The distributed hierarchical control plane architecture was used to reduce the delay of the controllers, and the QAR algorithm with the help of reinforcement learning could have good time-efficiency with the experiments. The method worked as described: the switch forwarded the first packet of the flow to the domain controller and requested the path. The current network state was updated based on the latest information from the slave controller via reinforcement learning method. The reinforcement learning offered the feasible path with the QoS requirements and modified the forwarding tables.

In 2020, Jiawei Wu et al. [11] proposed a deep reinforcement learning and DDPG-based cognitive routing, a quality-aware algorithm. The goal of the routing algorithm was to show feasibility and a network simulation utility was developed to aid further research of cognitive routing algorithms. The method makes use of the network diagnostic metrics to make routing decisions. The cognitive routing was used as an extension for the ability to infer the network state and learn policies based on it. The problem is modeled as a reinforcement learning (RL) task where the network is the environment, and the routing controller is the agent. Furthermore, the state is encapsulated as the amount of traffic (in total size) of data flowing between each pair of routers. The actions correspond to updating the weights which govern how a router chooses which interface to forward to. Reward is computed based on the delay, where lesser delay corresponds to greater reward. Q-learning with Deep Deterministic Policy Gradient (DDPG) is used to solve the RL task. Wu et al. simulated their algorithm using a newly developed utility, called RL4net. After simulating, it was found that both the critic and actor network were able to be successfully trained. The algorithm displayed a lower average end-to-end delay.

## 3.2   Decentralized Routing Algorithms

### 3.2.1 Supervised Learning Methods

In 2017, N. Kato et al. [5] proposed a deep learning routing algorithm which uses the history of network traffic to adaptively route packets. Each node was the unit for the neural networks. The input of the model was an array of numbers representing how many packets were forwarded through each node in the network, and the output was the interface to forward the packet along. The algorithm worked in three phases. First, OSPF was used to collect training data. Then, the network was trained using a greedy layer-wise approach, with back-propagation used to fine tune afterwards. Finally, the network used the trained networks to make routing decisions. An interesting optimization was that the networks trained at each node were sent to every edge node, and the path was decided at the node that receives the packet initially. The deep learning method outperformed the OSPF baseline on the experimental 16-node network.

In 2017, B. Mao et al. [12] introduced the routing strategy based on the deep learning architecture, Deep Belief Architectures (DBA). The input of the deep learning model was the traffic pattern observed at each router, and the pattern was defined as the relationship between the quantity of inbound packets and the time interval. The DBA consisted of multiple layers of Restricted Boltzmann Machines (RBM), which included one visible layer and one hidden layer. As for the training step, they first used the Greedy Layer-Wise method to train the model, and then the values of weights and biases were updated based on gradient descent method to maximize the value of the log-likelihood function. The experiment results showed the method they proposed outperformed the benchmark method on the performance of the delay, throughput, and signaling overhead.

In 2018, Geyer and Carle [13] proposed a method for automated protocol design for automated network control and management. The method was based on the extension of Graph Neural Network (GNN). The model for distributed routing protocols had two important aspects, including how to distribute topology information among different nodes, and how to compute routes given a topology and link weights. In this model, the interface of the router was regarded as the additional nodes in the graph and the hidden node representation storing to the messages from the neighbor nodes contributed to the destination router we wanted from the routing algorithm. The model used two algorithms to learn its routing, including shortest path and max-min routing. Based on the evaluation results, the approach to the distributed routing protocols performed good. The accuracy could reach 98% and 95% for the two routing methods, respectively.

In 2018, D. K. Sharma et al. [14] proposed MLProph, a machine learning enhanced version of a prior work, PROPHET+, a probabilistic routing approach for opportunistic networks (OppNets) in which link performance and connectivity is variable and intermittent. The authors used two alternatives to enhance PROPHET+, neural networks and decision trees. Each had access to a variety of input parameters such as buffer capacity, node popularity, and number of successful deliveries. A neural network was trained for each link, and the output represents whether successful delivery was likely if the packet is forwarded along this link. Backpropagation was used to train the networks against the sample data. The second alternative uses decision trees. Each decision node divided the sample data into two classes by comparing one of the inputs to a value, and leaf nodes represented a classification the same as in the neural network model. Both alternatives outperform PROPHET+ in both delivery chance and average latency.

In 2019, J Reis et al. [15] proposed a flexible supervised learning framework via learning paths from optimal Mixed Integer Linear Programming (MILP) and the trained model incorporating all the path pairs. Instead of always choosing the shortest path, the Deep Neural Networks method gave the routing path minimizing the congestion for the network system. The paper formulated the problem of finding the best path for a given sequence of flows by optimizing the function for minimizing the congestion of networks with maximum link utilization and the congestion measurement proposed by Fortz and Thorup [16].

There were two parts in the supervised-learning-based (SL-based) routing system, including the deep neural network (DNN) and the post-processing routing algorithm. The DNN was used to determine the path, and the post-processing model was used to filter the invalid path. The SL-model took the concatenation of the features of the flow, including source name, destination name and bandwidth as input, and output the optimal path according to the congestion cost for each flow. The dataset used for the model was based on topology and TMs from GEANT network, and the same information from the Abilene network. The results of the model greatly outperformed the shortest path method for both MLU and Fortz metrics.

### 3.2.2 Reinforcement Learning Methods

In 1993, Justin Boyan et al. [17] proposed an adaptive routing policy based on Q-learning, called Q-routing. Each node had its own Q function to learn. The Q function took the destination and candidate next hop node as inputs and outputted the estimated time to deliver the packet using the next hop. This time included any queuing and transmission delays. When a packet is forwarded from node X to Y, Y responded to X with Y's estimated time of delivery for the packet. Using this X could in turn update its Q function to account for any differences in estimated time. (It is similar to distance vector algorithm with estimated times instead of link costs). The Q function was stored as a 2-D table with a nodes' neighbors as one dimension and all possible destinations as the other and the estimated time as the value. Compared to the shortest path, Q-routing performed roughly the same under low network load but outperformed under high load by adapting to avoid congestion. Q-routing was also capable of adapting to changing network conditions and topology. The equation for updating the Q value is shown in the Eq. (1), where $Q_x(y,d)$ represents the time that node $x$ estimates it takes to deliver a packet bound for node $d$ by way of $x$'s neighbor node $y$, $q$ means the packet takes $q$ steps in $x$'s queue, $s$ means the steps needed for transmission from $x$ to $y$, and $\eta$ is the learning rate [17].

$$\Delta Q_x(y,d) = \eta\big(min_{z \in N(y)} Q_y(z,d) + q_x + \delta - Q_x(y,d)\big) \qquad (1)$$

In 1998, Shailesh Kumar et al. [18] proposed the CQ-Routing algorithm, which used the confidence values to improve the quality of exploration for adaptive packet routing in communication networks. Instead of treating the learning rate as a constant, CQ-Routing uses a confidence value for each Q value to determine the learning rate for each update. The experiments showed that CQ-Routing outperformed OSPF, BF, and Q-Routing on average delivery time for any kind of load and was more adaptive to the network topology changes. The equations for updating C value and Q value are shown in the Eq. (2), Eq. (3),and Eq. (4), where $Q_x(y,d)$ represents the time that node $x$ estimates it takes to deliver a packet bound for node $d$ by way of $x$'s neighbor node $y$, $q$ means the packet takes $q$ steps in $x$'s queue, $s$ means the steps needed for transmission from $x$ to $y$, and $\lambda$ is the constant decay rate[18]. When Q values are not updated in last step, we use Eq. (3a) to calculate $C$, otherwise, we use Eq. (3b).

$$\eta(C_{old}, C_{est}) = max(C_{est}, 1 - C_{old}) \qquad (2)$$

$$C_{new} = \begin{cases} \lambda C_{old} & (3a) \\ C_{old} + \eta(C_{old}, C_{est})(C_{est} - C_{old}) & (3b) \end{cases}$$

$$\Delta Q_x(y,d) = \eta(C_{old}, C_{est})\big(min_{z \in N(y)} Q_y(z,d) + q_x + \delta - Q_x(y,d)\big) \qquad (4)$$

In 2002, Will Newton et al. [19] proposed the NNQ-Routing algorithm, which used a neural network approximation to help improve the scalability of Q-Routing. The original Q-learning based Q-Routing algorithm used a table store the Q function, but this did not scale up. In the paper, the Q function was replaced by a 3-layer perceptron model, where the bits of the source and destination IP were the input and the output is a linear node intended to approximate the delay for using the link. Delays and dropped packets (caused by excessive forwarding) were penalties while successful deliveries are rewards in the Q-learning method. Two training alternatives were used: traditional Q-learning off policy was used as well as Sarsa method on policy. The experiment was an overall failure with each configuration either diverging or not showing any sense of converging in the best case. It was believed the non-linear nature of the neural network caused the Q function to fail to converge.

In 2009, B. Xia et al. [20] proposed the spectrum-aware DRQ-routing protocol for routing in multi-hop cognitive radio networks (CRN). The method included two adaptive reinforcement learning techniques, Q-Learning and Dual Reinforcement Learning. The backward exploration and forward exploration are used in the model. The Q-value routing table was used to store the cognitive nodes to estimate the quantity of the available routes. The method was more cost-effective, and more sensitive to the dynamic routing and avoided the limitations in on-demand protocols. The experiments showed that their protocol outperformed the shortest path methods on the metrics of average packet delivery time in high load condition (>2.5 packets/s).

### 3.3 Summary

One of the advantages of the centralized routing algorithms is that they can save the resource. Just one table is built and control all the network routing. Also, it is easy to modify the routing policies based on the control plane, which makes the centralized routing algorithm flexible. Besides, since the centralized routing algorithms based on the full network topology and possibly network state information, they can take full consideration in a global view. A limitation depending on the method is needed to communicate with the central controller for routing decisions. Therefore, when changes happen in the forwarding table, it needs more time and more capacity to updates. Besides, when it comes the attack or problem in the control plane, it may have the single point of failure.

In decentralized algorithms, there is no central controller, so some mechanism must be used to allow access to further network state. Each of the node will store a forwarding table, and will decide the routing policy for itself, which means it is more scalable since you can easily modify some parts of the routing methods by yourself. The methods can adaptively react to the changes well. But at the same time, the decentralized methods require more space and there may infinity loop problems.

In most supervised learning methods, the inputs, outputs of the methods, hyperparameters and how to configure the embeddings must be carefully constructed. But once we have the good preprocessing, the model is very scalable. However, in supervised learning methods, training dataset need to be generated from simulations to make the model work well in realistic scenarios.

A benefit for reinforcement learning is that the methods are often trained on-line, reacting to rewards from the environment. They can be continually learning. However, carefully constructing the rewards and responses function, like the Q function in Q-learning, is still important. As can see in [19], the neural network Q function approximator was not sufficient to model the Q function. That is why lots of variant Q-Routing algorithm come out.

As for the cons and pros for each of the 14 methods, we have made a comparison table shown in Table. 1.

| Method | Type | Machine Learning Technique | Pros | Cons |
|---|---|---|---|---|
| GADL [6] | Centralized Supervised | Graph Neural Networks | Training faster; High accuracy; Reduce average latency | Data hungry |
| ML-assisted LL [7] | Centralized Supervised | Naive Bayes Classifier | Reduce blocking chance | More experiments for circuit-switched network; Data hungry |
| RouteNet [8-9] | Centralized Supervised | Graph Neural Networks | Minimize the cost for getting performance for different routing policies; Reduce the average delay and packet loss | Complicated preprocess |
| QAR [10] | Centralized Reinforcement Learning | SARSA Reinforcement Learning | Fast convergence; Minimize the signaling delay | Need to improve the scalability |
| DRL-based Cognitive Routing [11] | Centralized Reinforcement Learning | Deep Reinforcement Learning, DDPG framework | Reduce average end-to-end delay | Limited experimental results; Extending to real-world network |
| DL Heterogeneous Traffic Control [5] | Decentralized Supervised | Deep Neural Network | Reduce average delay and increase the throughput | Extend to real-world network |
| DBA Routing [12] | Decentralized Supervised | Deep Belief Architecture | Reduce average delay and increase the throughput | Need to improve the scalability |
| GQNN Routing [13] | Decentralized Supervised | Graph-Query Neural Network. | Generated Models with good delay | Extend to other routing protocols |
| MLProph [14] | Decentralized Supervised | Neural Network, Decision Tree | Reduce average delay and packet loss chance | Simulate with real mobility traces; Test other novel ML models |
| DNN Routing [15] | Decentralized Supervised | Deep NN, MILP | Improved link utilization and reduce the congestion | Data hungry |
| Q-Routing [17] | Decentralized Reinforcement Learning | Q-Learning | Improved performance in high load conditions | Q function table does not scale well to larger networks |
| CQ-Routing [18] | Decentralized Reinforcement Learning | Q-Learning, Confidence Learning | Faster, and sustain higher network load | Extend to real world network |
| NN Q-Routing [19] | Decentralized Reinforcement Learning | Neural Network, Q-Learning | Improved scalability of Q function | Not scale with some network case |

| DRQ Routing[20] | Decentralized Reinforcement Learning | Dual Reinforcement Learning, Q-Learning | Improved delivery time under high load conditions | Extend to real-world network |

Table 1: Comparison between different machine learning application in routing

## 4 Performance comparison

We implement basic Q-routing method [18] and the CQ-routing method [19] based on the papers and resources for Q-Routing [21]. After understanding, we implemented the code, and followed the paper [18] and [19] to set up the experiment for each of these two models to compare the average delivery time of these two models with different network load. The interpretation of the experiment is shown in 4.1, and the results and discussions are described in 4.2.

### 4.1 Simulation/Evaluation Settings

We use a simulated network with a set of nodes and links in the experiment. Within the experiments, random nodes in the network will receive the packets with a random destination periodically. We treat the number of the packets generated per simulation step as the network load. If in some step, multiple packets are generated, we use an unbounded FIFO queue to store them, and each step, only the first packets in the queue can be processed. Within the experiments, we used 2.15 packets/step as the medium loads, and 2.75 packets/step as the high loads.

| Medium Load | High Load |
| --- | --- |
| 2.15 packets/step | 2.75 packets/step |

Table 2: Loads settings

We define the delivery time of one packet as the time between it occurs in the source node and disappears in the destination node. And the measurement for the delivery time is based on the simulation steps. For example, if it takes 10 loops for packet A to reach the destination node, then the delivery time is 10 simulation steps.

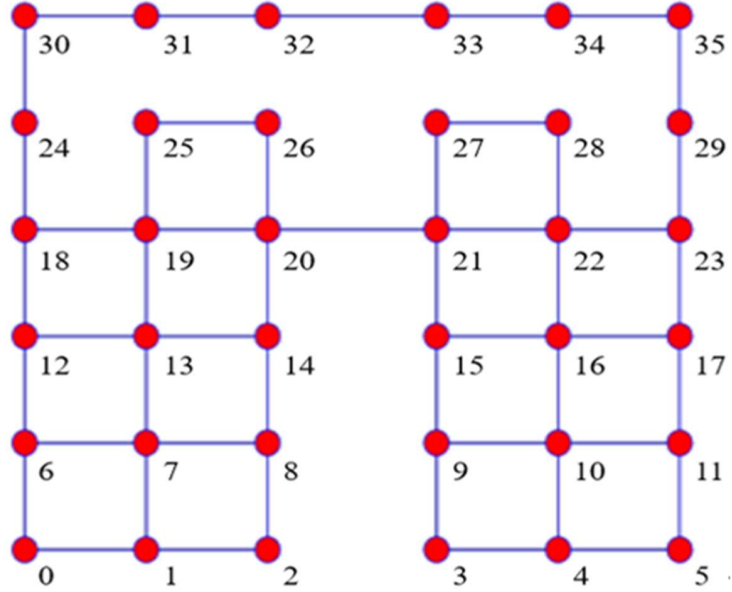

Figure 2 The irregular 6x6 grid topology[18]

The experiments are carried out between CQ-Routing and Q-Routing on the irregular 6x6 grid topologies proposed by Boyan and Littman in 1994[18] shown in Fig. 1. In the 6x6 irregular grid, there are only two types of route to support the packet transmitted from the nodes in left side, including {0,1,2, 6,7,8,12,13,14,18,19,20,24,25,26,30,31,32}, to the nodes in the right side, including {3,4,5,9,10,11,15,16,17,21,22,23,27,28,29,33,34,35}. One of them is that the link (20,21) is in the route, and the other one is that the link (32,33) is in the route.

The results are the comparison of the average delivery time under different loads between CQ-Routing and Q-Routing. The average here is the average on 10 tests with random starts. Follow the paper [17] and [18], we choose the best hyperparameters the learning rate $\eta = 0.85$ for Q-Routing, and the decay constant $\lambda = 0.95$ for CQ-Routing.

## 4.2 Results and Discussions

Based on Eq.(1), Eq.(2), Eq.(3) and Eq.(4), we implement the codes. And based on the experiment settings above. Here are the results for low load, medium load, and high load for Q-Routing, CQ-Routing.

The relative performance is shown in Fig. 3 and Fig. 4. At medium load levels (2.15 packets/step), CQ-Routing learns the routing policy faster than Q-Routing method as shown in Fig. 3. At high load levels (2.7 packets/step), CQ-Routing learns the routing policy faster than Q-Routing method as shown in Fig. 4. The results satisfy the conclusion for CQ-Routing and Q-Routing mentioned before in section 3.

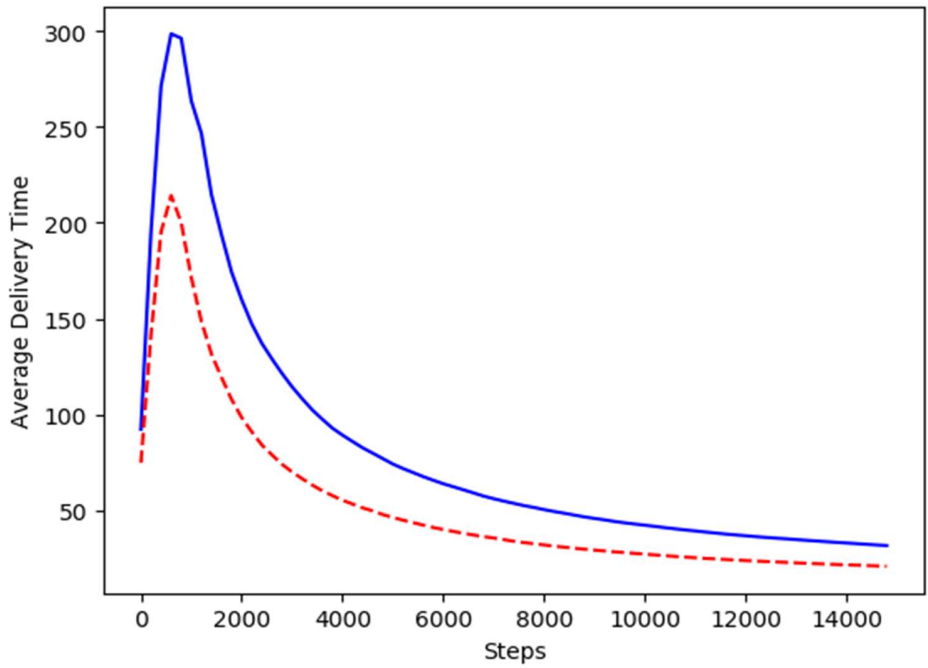

Figure 3: Learning at medium network load; red dot line represents the CQ-Routing, and blue line represents the Q-Routing

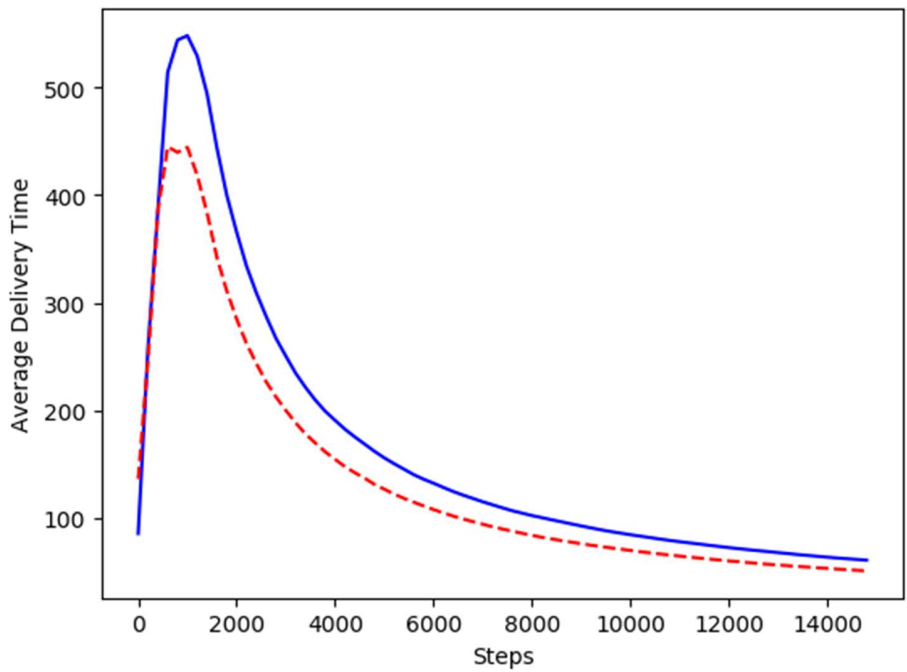

Figure 4: Learning at high network load; red dot line represents the CQ-Routing, and blue line represents the Q-Routing

## 5    Conclusion

Development of new routing algorithms is of clear importance as the volume of Internet traffic continues to increase. As this survey there is much research into how Machine Learning techniques can be employed to improve the performance and scalability of routing algorithms. We surveyed both centralized and decentralized ML routing architectures and using a variety of ML techniques broadly divided into supervised learning and reinforcement learning. Many of the papers showed promise in their ability to optimize some aspect of network routing. We also implemented two routing protocols and verified the efficacy of their results.

While the results of most of the papers showed promise, many of them are based on simulations of potentially unrealistic network configurations. To provide further efficacy to the results, more real-world results are necessary. As the prevalence of SDN grows, its flexibility may provide the framework to support this data. We may also see more network metrics being used as input to the ML techniques, as well as more experimentation with other ML techniques. Besides, there is little research based on combining reinforcement learning and supervised learning, which can be definitely be one of the future directions.

## 6    Contribution

Ke is elected as the team leader, and Mitchel (50%) and Ke (50%) both worked on all the parts of the work together. More specific, we both wrote the idea of each part, and discussed how to combine them together.